
\input harvmac
%
%
%
%
\ifx\answ\bigans
\else
\output={
  \almostshipout{\leftline{\vbox{\pagebody\makefootline}}}\advancepageno
}
\fi
%
%
%
\def\mayer{\vbox{\sl\centerline{Department of Physics 0319}%
\centerline{University of California, San Diego}
\centerline{9500 Gilman Drive}
\centerline{La Jolla, CA 92093-0319}}}
%
%

%
%
%
%
\def\abstract#1{\centerline{\bf Abstract}\nobreak\medskip\nobreak\par #1}
%
%
%
%
\edef\tfontsize{ scaled\magstep3}
 \tfontsize  \tfontsize
 \tfontsize \font\titlei=cmmi10 \tfontsize
\font\titleis=cmmi7 \tfontsize \font\titleiss=cmmi5 \tfontsize
\font\titlesy=cmsy10 \tfontsize \font\titlesys=cmsy7 \tfontsize
\font\titlesyss=cmsy5 \tfontsize  \tfontsize
\skewchar\titlei='177 \skewchar\titleis='177 \skewchar\titleiss='177
\skewchar\titlesy='60 \skewchar\titlesys='60 \skewchar\titlesyss='60
%
%
%
%
%
\def\inv{^{\raise.15ex\hbox{${\scriptscriptstyle -}$}\kern-.05em 1}}
\def\lbar{{\lower.35ex\hbox{$\mathchar'26$}\mkern-10mu\lambda}} 

%
%
%
%
\def\dsl{\,\raise.15ex\hbox{/}\mkern-13.5mu D} 
\def\delsl{\raise.15ex\hbox{/}\kern-.57em\partial}
\def\Ksl{\hbox{/\kern-.6000em\rm K}}
\def\Asl{\hbox{/\kern-.6500em \rm A}}
\def\Dsl{\hbox{/\kern-.6000em\rm D}} 
\def\Qsl{\hbox{/\kern-.6000em\rm Q}}
\def\gradsl{\hbox{/\kern-.6500em$\nabla$}}
%
%
\def\lspace{\ifx\answ\bigans{}\else\qquad\fi}
\def\lbspace{\ifx\answ\bigans{}\else\hskip-.2in\fi} 
%
%
\def\boxeqn#1{\vcenter{\vbox{\hrule\hbox{\vrule\kern3pt\vbox{\kern3pt
        \hbox{${\displaystyle #1}$}\kern3pt}\kern3pt\vrule}\hrule}}}
%
%
\def\mbox#1#2{\vcenter{\hrule \hbox{\vrule height#2in
\kern#1in \vrule} \hrule}}
%
%
%
%

   \def\CL{{\cal L}}
  \def\CO{{\cal O}}

%
%
%
%
%

%

\def\bar#1{\overline{#1}}

\def\lform{\hbox{$\sqcup$}\llap{\hbox{$\sqcap$}}}
\def\darr#1{\raise1.5ex\hbox{$\leftrightarrow$}\mkern-16.5mu #1}

%
%
\def\frac#1#2{{\textstyle{#1\over #2}}} 
%
%
%
%

\def\Tr{\mathop{\rm Tr}}

%
%
%
%

%
%
\def\ltap{\ \raise.3ex\hbox{$<$\kern-.75em\lower1ex\hbox{$\sim$}}\ }
\def\gtap{\ \raise.3ex\hbox{$>$\kern-.75em\lower1ex\hbox{$\sim$}}\ }
\def\gl{\ \raise.5ex\hbox{$>$}\kern-.8em\lower.5ex\hbox{$<$}\ }
\def\roughly#1{\raise.3ex\hbox{$#1$\kern-.75em\lower1ex\hbox{$\sim$}}}
%
%

%

%
\def\np#1#2#3{{Nucl. Phys. } B{#1} (#2) #3}
\def\pl#1#2#3{{Phys. Lett. } {#1}B (#2) #3}
\def\prl#1#2#3{{Phys. Rev. Lett. } {#1} (#2) #3}
\def\physrev#1#2#3{{Phys. Rev. } {#1} (#2) #3}

\relax

\def\CO{{\cal O}}

\def\lta{\ \hbox{\raise.55ex\hbox{$<$}} \!\!\!\!\!
\hbox{\raise-.5ex\hbox{$\sim$}}\ }
\def\gta{\ \hbox{\raise.55ex\hbox{$>$}} \!\!\!\!\!
\hbox{\raise-.5ex\hbox{$\sim$}}\ }
\def\mayer{\vbox{\sl\centerline{Department of Physics}
\centerline{9500 Gilman Drive 0319}
\centerline{University of California, San Diego}
\centerline{La Jolla, CA 92093-0319}}}

\def\mayer{\vbox{\sl\centerline{Department of Physics}
\centerline{9500 Gilman Drive 0319}
\centerline{University of California, San Diego}
\centerline{La Jolla, CA 92093-0319}}}
\def\[{\left[}
\def\]{\right]}
\def\({\left(}
\def\){\right)}
\def\toright#1{\smash{\mathop{\longrightarrow}\limits_{#1}}}
\def\psl{\hbox{/\kern-.6000em $p$}}
\def\frac#1#2{{\textstyle{#1 \over #2}}}
\noblackbox
\vskip 1.in
\centerline{{\titlefont{A Method for Simulating  }}}
\vskip .15in
\centerline{{\titlefont{ Chiral Fermions on the Lattice}}}
\bigskip
\centerline{{\bf Submitted to Physics Letters B}}
\vskip .5in
\centerline{
David B. Kaplan\footnote{$^{\dagger}$}
{Sloan Fellow, NSF Presidential
Young Investigator, and DOE Outstanding}
\footnote{}{Junior
Investigator. E-mail: dkaplan@ucsd.bitnet}
}
\bigskip\medskip
\mayer
\bigskip\bigskip
\abstract{I show that a lattice theory of massive interacting fermions in
$2n+1$ dimensions may be used to simulate the behaviour of
massless chiral fermions in $2n$ dimensions if the fermion mass has
a step function shape in the extra dimension.  The massless states arise
as zeromodes bound to the mass defect, and all doublers can be given
large gauge invariant masses. The manner in which the
anomalies are realized is
transparent: apparent chiral anomalies in the $2n$ dimensional subspace
correspond to charge flow into the extra dimension.
   }
\vfill
\hbox{\hbox{ UCSD/PTH 92-16\hskip 2in May 1992} }
\eject

A successful  chiral regulator for gauge theories has
proven elusive.   Dimensional regularization explicitly violates the
chiral gauge symmetry, as does  the conventional Pauli-Villars
technique; zeta function regularization
fails because the chiral Dirac operator does not provide an eigenvalue
problem. A lattice regularization technique would be desirable, allowing
one to simulate the standard model as well as strongly coupled chiral
gauge theories \ref\scgt{See, for example, H. Georgi, \np{156}{126}; S.
Dimopoulos, S. Raby and L. Susskind, \np{173}{1980}{208}}. However, the naive
approach suffers from the existence of doublers that render the theory
vector-like instead of chiral \ref\ksm{L.H. Karsten and J. Smit,
\np{183}{1981}{103}}. Some more sophisticated approaches appear to suffer
other  obstructions \nref\bod{G.T. Bodwin and E.V. Kovacs,
\physrev{D35}{1987}{3198}; {\it ibid.} D37 (1988) 1008; {\it ibid.} D41
(1990) 2026}\nref\petcher{M.F.L.Goltermann, D.N.Petcher, J.Smit,
\np{370}{1992}{51}}\nref\ep{M.F.L.Golterman, D.N.
Petcher, {\it On the Eichten-Preskill Proposal...}, Washington
University preprint, WASHU-HEP-91-61.}\nref\banks{T. Banks,  Phys.Lett.
272B,(1991),75; T. Banks and A. Dabholkar, {\it Decoupling a Massive
Fermion...}, Rutgers preprint RU-92-09, hep-lat/9204017.}\nref\dugan{
M. J. Dugan, L. Randall, {\it On The Decoupling of Doubler Fermions...}
MIT preprint, MIT-CTP-2050, {\it Submitted to
Nucl.Phys.B}.}\refs{\bod-\dugan}, although some have yet to be
thoroughly tested  \ref\extant{L. Maiani, G.C. Rossi, M. Testa,
\pl{261}{1991}{479}; A. Borrelli, L. Maiani, R. Sisto, G.C.
Rossi and M. Testa, \np{333}{1990}{335}; M. Bochicchio, L.
Maiani, G. Martinelli, G. C. Rossi and M. Testa,
\np{262}{1985}{331}; G.T. Bodwin and E.V. Kovacs, Nucl.
Phys. B (Proc. Suppl.) 20 (1991) 546; C. Pryor,
\physrev{D43}{1991}{2669}}. Any such regulator must
correctly reproduce  the anomalous Ward identities of the continuum
theory. Taking the continuum limit must fail  in theories where
the  gauge currents  have anomalous divergences; on the other hand,
nonconservation of global  currents must be allowed in the continuum.
As Banks has emphasized recently, a necessary criterion for chiral
lattice theories is that   they must correctly
describe  proton decay in a simulation of the standard model \banks.
The point is that the standard model has anomaly free chiral gauge
currents, but anomalous global currents, such as baryon number
\ref\thooft{G. `t Hooft, \prl{37}{1976}{8};
\physrev{D14}{1976}{3432}}. A lattice regulator that achieves this
while preserving gauge invariance will have to be devious, as ``no-go''
theorems imply \ref\nn{H.B. Nielsen and M. Ninomiya, \np{185}{1981}{20};
\np{195}{1982}{541E}; \np{193}{1981}{173}; \pl{105}{1981}{219}; L.H.
Karsten, \pl{104}{1981}{315}}.

In this Letter I show that chiral fermions in $2n$ dimensions
may be simulated by  Dirac fermions in $2n+1$
dimensions with a space dependent mass term.  In particular, when
 there is a domain wall defect in the mass parameter,  the lattice theory
will exhibit massless chiral zeromodes bound to the $2n$ dimensional
domain wall, as well
as doublers with the opposite chirality.  Since the original $2n+1$
dimensional theory is vector-like, the doublers may be removed by means
of a gauge invariant Wilson term.  This system can be used to simulate a $2n$
dimensional chiral gauge theory (for $n=1,2$) provided that the massive
fermion modes
which are not bound to the mass defect decouple, and  provided that the
gauge fields can be constrained to propagate along the defect.  I argue
that whether or not these conditions can be met  depends on an algebraic
relation which is just the one that ensures that the $2n$ dimensional
gauge currents are exactly divergenceless. Global currents,
however, may have anomalous divergences even in the regulated theory
before taking the continuum limit, due to the curious role played by the
extra dimension.

I begin by recapitulating some well known facts about chiral zero modes
of a free Dirac fermion  in $2n+1$ dimensions coupled to a domain wall.
I work in Euclidian space and take the $2n+1$
coordinates to be $z_\mu=\{\vec x, s\}$ where $\mu$ runs over
$1,\dots,2n+1$ while $\vec x$ is the coordinate in the $2n$ dimensional
subspace.  I use   hermitian   gamma matrices satisfying
$\{\gamma_\mu,\gamma_\nu\} = 2\delta_{\mu\nu}$
 and define
\eqn\gam{
\gamma_{2n+1}\equiv \Gamma_5 \ .}
Note that $\Gamma_5$ is the chiral operator in the $2n$ dimensional
theory.  The fermion mass is assumed to have the form of a domain
wall---a monotonic function of $s$ with the asymptotic form
\eqn\mass{m(s)\toright{s\to\pm\infty} \pm m_{\pm},\qquad
 m_{\pm}>0\ .}
The Dirac operator is then given by
\eqn\cdop{\hat K_D = [\delsl + m(s)] = [\vec\gamma\cdot\vec \nabla +
\Gamma_5\partial_s + m(s)]\ .}
Since the mass $m(s)$ vanishes at $s=0$ one might expect
massless states bound to the defect. Such ``zeromodes'' would
have to satisfy $\hat K_D \Psi_0 = \vec\gamma\cdot\vec\nabla \Psi_0$ in
order to describle massless propagation along the defect.  Specifically,
the zeromodes must have the form
\eqn\zmsol{
\Psi_0^{\pm} = e^{i\vec p\cdot \vec x} \phi_{\pm}(s)u_{\pm}}
where $u_{\pm}$ are constant $2^n$ component chiral spinors
$$\Gamma_5 u_{\pm} = \pm u_{\pm}$$
and the functions $\phi_{\pm}(s)$ satisfy
\eqn\phisol{[\pm\partial_s + m(s)]\phi_{\pm}(s) = 0\ .}
The solutions to the above equation are
\eqn\psip{\phi_{\pm}(s) = \exp\left(\mp\int_0^{s} m(s'){\rm
d}s'\right)\ ,}
however with the asymptotic form for $m(s)$ given in eq. \mass,
only $\phi_+(s)$ is normalizable.
Thus there is a single, positive chirality massless fermion bound to the
mass defect.

What makes this model of interest for a lattice approach to chiral
fermions is that the $2n+1$ dimensional theory has neither chirality nor
anomalies, yet possesses a low energy effective theory (well below the
asymptotic mass scales $m_{\pm}$) which is a theory of a massless chiral
fermion in $2n$ dimensions.  As I will discuss below, the anomalies in
the effective $2n$ dimensional theory are realized through finite
Feynman diagrams and therefore have their canonical form  even in the
regularized theory.

A similar model can be formulated on a $2n+1$ dimensional lattice with
lattice spacing $a$ and sites labelled by $z=n_z a$.
The action is given by
\eqn\dact{S_D = \sum_z\ \bar\Psi_z \hat K_D \Psi_z}
where the lattice Dirac operator is given by
\eqn\latdir{ \hat K_D \Psi_z =
{1\over2a} \sum_{\mu=1}^{2n+1}\  \gamma_\mu (\Psi_{z+\hat\mu} -
\Psi_{z-\hat\mu})
 + m(s) \Psi_z}
and $\hat\mu$ corresponds to a displacement by $a$ in the $z_\mu$ direction.
For simplicity I take the fermion mass to be a
 step function
\eqn\latmass{m(s) =  m_0\theta(s)\equiv
{\sinh(a \mu_0)\over a}\theta(s)\ ,\qquad
\theta(s)= \cases{
-1, &$s\le -a$\cr
\ \ 0, &$s=0$\cr
+1, &$s\ge +a$\cr}\ .}
Then there are two chiral zeromode solutions $\Psi^{\pm}_0$ given by
\eqn\czm{
 \Psi_0^{\pm}(\vec p\,, z) = e^{i\vec p\cdot\vec x} \phi_{\pm}(s,\vec p\,)
u_{\pm}}
where the transverse wavefunctions are given by
\eqn\twf{\eqalign{
\phi_+(s,\vec p\,) &= e^{-\mu_0\vert s\vert}\cr
\phi_-(s,\vec p\,) &= (-1)^{n_s}\phi_+(s,\vec p\,)\  .}}
These solutions satisfy
\eqn\eig{
\hat K_D \Psi^{\pm}_0(\vec p\,,z)  = (i/a)\vec \gamma\cdot\sin (\vec p\,
a) \Psi^+_0(\vec p\,,z)\ ,\qquad \Gamma_5 \Psi^{\pm}_0 = \pm\Psi^{\pm}_0}
and, unlike the continuum example, are both normalizable and localized
along the mass defect at $s=0$.
For $\vert\vec p a\vert\ll 1$, $i\vec \gamma\cdot \sin(\vec p\, a)/a \to
i\psl$,
the inverse propagator for a massless mode travelling in $2n$
dimensions.  Therefore  the mode $\Psi^+_0$  for small $\vec p$
corresponds to the positive chirality state \zmsol\ found in the
continuum.  However,
 the lattice model is seen to actually describe $2^{2n+1}$ massless modes in
the continuum, rather than one, corresponding to $\Psi^+_0$ and
$\Psi^-_0$  with $\vec p$ near the $2^{2n}$ corners of the $2n$
dimensional Brillouin zone.
The  conventional analysis  of doubler modes \ksm\ reveals that these
$2^{2n+1}$ modes correspond to $2^n$ positive chirality states and $2^n$
negative chirality states.  The resultant lattice spectrum is therefore
completely
vectorlike in the continuum, unlike what I wish to describe.

Nevertheless, since the $2n+1$ dimensional theory \dact\ is itself vectorlike,
it is possible to add a Wilson term to eliminate the unwanted doublers.
Thus I add to $S_D$ the action
\eqn\wact{\eqalign{
S_W &= a w \sum_z \bar \Psi_z \hat K_W \Psi_z\ ,\cr
\hat K_W \Psi_z &= \sum_{\mu=1}^{2n+1} \Delta_\mu
\Psi_z=\sum_{\mu=1}^{2n+1} ( \Psi_{z+\hat\mu}
-2\Psi_z + \Psi_{z-\hat\mu})\ .}}
Zeromodes will now correspond to solutions satisfying eq. \eig\ with
$\hat K_D$ replaced by $\hat K_D + aw\hat K_W$.  Thus the transverse
wavefunctions $\phi_{\pm}$ must satisfy
\eqn\twfe{\left[\pm \partial _s +  m_0\theta(s) +wa\Delta_s +
(2w/a)\sum_{i=1}^{2n}(\cos( p_i a)-1)\right] \phi_{\pm}(s) = 0\ .}
I will analyze this equation for the particularly convenient choice of
 $w=.5 $, though there is actually a broad range of
values that will have a similar solutions.  Defining
$$F(\vec p\,) = \sum_{i=1}^{2n}(1-\cos ap_i)\ge 0\ ,$$
eq. \twfe\ becomes (for $w=.5$)
\eqn\xxx{\eqalign{ \phi_{\pm} (s\pm a) &= -m_{\rm eff}(s)\phi_{\pm}(s)\
,\cr
m_{\rm eff}(s)= am_0\theta(s)-1&-F(\vec p\,)  = \cases{ +am_0-F(\vec p\,)- 1
&$s>0$\cr -am_0-F(\vec p\,)-1 & $s<0$}}}
In order for there to exist {\it normalizable} solutions for
$\phi_{\pm}(s,\vec p\,)$,  $\vert m_{\rm eff}\vert$ must satisfy the
following conditions:
\eqn\meff{\eqalign{
&\phi_+:\ \qquad  \vert m_{\rm eff}\vert > 1\ {\rm for\ }
s<0,\ \ \vert m_{\rm eff}\vert < 1\ {\rm for\ }
s>0;\cr
&\phi_-:\ \qquad  \vert m_{\rm eff}\vert < 1\ {\rm for\ }
s<0,\ \ \vert m_{\rm eff}\vert > 1\ {\rm for\ }
s>0\ .}}
Since $am_0$ and $F(\vec p\,)$ are positive, the conditions for a
normalizable $\phi_-$ solution can never be met,
and so all $2n$ of the $\Psi^-_0$ modes have been
eliminated.  On the other hand, there are normalizable $\phi_+(s,\vec
p\,)$ solutions for all $\vec p$ satisfying
\eqn\cond{0< am_0-F(\vec p\,)<2\ .}
Thus, for $0<am_0<2$ there is always a normalizable solution for small
$\vec p$, since $F(0)=0$.  Furthermore, on the Brillouin boundary $F\ge
2$, and so all of the doubler solutions for $\phi_+$ have been removed
for the same range of $am_0$. This
shows that the Wilson action \wact\ has eliminated all but one of the
$2^{2n+1}$ zeromodes bound to the domain wall in the continuum limit.
Only a single positive chirality mode remains.

It may seem surprising that
the Wilson term could remove an odd number of chiral modes, since one
expects them to be removed in pairs with opposite chirality.  In this case
though, the chiral doubler modes are paired with modes ``at infinity'',
rather than with each other.  The effect is more easily understood in
finite volume. Consider a lattice of length $L$ in the $s$ direction,
with periodic boundary conditions.  Now if there is a domain wall at
$s=0$, there is necessarily an ``anti-domain wall'' at $s=\pm L/2$.  For
every zeromode at $s=0$, there is a zeromode of opposite chirality at
$s=\pm L/2$. The vectorlike nature of the $2n+1$ dimensional theory is
now manifest: every mode is chirally paired, though the $s$ dependent
mass function has allowed one to separate some chiral pairs by a distance
$L/2$ with exponentially small overlap.

At finite volume one can also see how the
peculiar discontinuity in the spectrum
implied by eq. \cond\ comes about.  At finite $L$ with periodic boundary
conditions, only the $\vec p=0$
mode is exactly chiral; for $\vec p> 0$ satisfying the condition \cond,
the modes have  masses exponentially small in $L$, corresponding to
a finite probability that the mode can tunnel to the anti-domain wall
and flip chirality.  As $L\to \infty$, this probability goes rapidly to
zero and the modes are exactly chiral as seen above. For values of $\vec
p$ not satisfying eq. \cond\ the modes are massive for any $L$, finite
or infinite.

I conclude that at infinite volume
the spectrum of the free theory  contains
\hbox{(i) doublers} with mass $\CO(w/a)$;
\hbox{(ii)} Dirac fermions with mass $\CO(m_0)$ that propagate in the
full  $2n+1$ dimensions;
\hbox{(iii)} an exactly massless chiral zeromode bound with
$\Gamma_5=+1$ bound to the $2n$ dimensional
defect.  For momentum $\vert\vec p\vert \ll 1/a$ the mode has the
conventional continuum propagator.
The theory is vector-like at every step: both the
doublers with mass at the cutoff, as well as the mass $m_0$ scattering
states, are all Dirac fermions.  The zeromodes are also paired: in
finite volume one sees that there are zeromodes of opposite
chirality bound to an anti-domain wall, which are carried off to spatial
infinity in the infinite volume limit.

When background gauge fields are coupled to the $2n+1$ dimensional Dirac
theory, there appears to be a paradox.
One would expect that at energies well below $m_0$ all massive  fermion
degrees would decouple, yielding an effective $2n$ dimensional theory of
zeromodes at one of the domain walls.  However, such a theory is in
general anomalous and cannot be simultaneously gauge invariant and
chirally invariant.  In the continuum, the zeromode gauge current
$J^{zm}$ should have a divergence equal to
\nref\anm{S. Adler, \physrev{177}{1969}{2426}; J.S. Bell and R. Jackiw,
Nuovo Cimento 60 (1969) 47}\nref\hdim{
 P.H. Frampton and T.W. Kephart,
\prl{50}{1983}{1343,1347}; {\it ibid.} 51 (1983) 232E;
\physrev{D28}{1983}{1010}; L. Alvarez-Gaum\'e and E. Witten,
\np{234}{1983}{269};
 B. Zumino, Lectures at Les Houches Summer School (1983); R.
Stora, Carg\`ese lectures (1983)}\refs{\anm-\hdim}
\eqn\anom{ D_i J^{zm}_i = i \Gamma_5 C_n \epsilon_{i_1\ldots i_{2n}}
F_{i_1 i_2}\cdots F_{i_{2n-1}i_{2n}}\ ,}
where
$${C_n = (-1)^n/(2^{2n}\pi^nn!)\ ,}$$
$\Gamma_5=\pm 1$ is the chirality of the zeromode, and the indices run
over the $2n$ dimensions.  On the other hand, the lattice theory is
manifestly finite and gauge invariant.  So what is going on?

The resolution of the paradox is that the massive fermion states that
live off the domain walls do not entirely decouple---even in the low
energy limit---and so the effective theory is not truly $2n$
dimensional.  Callan and Harvey first described this effect some years
ago \ref\ch{C.G. Callan, Jr., and J.A. Harvey,
\np{250}{1985}{427}}
while  studying the properties of zeromodes on
topological defects and their relation to the anomaly descent relations
\nref\desa{B.
Zumino, Y.-S. Wu and A. Zee, \np{239}{1984}{477}}\nref\desb{L.
Alvarez-Gaum\'e and P. Ginsparg, Ann.
Phys. 161 (1985) 423; {\it ibid.} 171, 233E; O. Alvarez, I. Singer and
B. Zumino, Commun. Math. Phys. 96 (1984) 409}\refs{\desa-\desb}.
They showed that the combined effect of the background field strength
$F_{\alpha\beta}$  and the space dependent mass term
\eqn\lmass{\CL_m = + m(s) \bar \Psi \Psi}
was to distort the heavy fermion vacuum,
producing a Goldstone-Wilczek current \ref\gwc{J.
Goldstone and F. Wilczek, \prl {47}{1981}{986}}\ far from the domain wall,
given by
\eqn\gwc{\langle \bar \psi \gamma_\mu \psi\rangle_{GW} =
-{i\over2}{m(s)\over\vert m(s)\vert} C_n
\epsilon_{\mu\alpha_1\ldots \alpha_{2n}}
F_{\alpha_1 \alpha_2}\cdots F_{\alpha_{2n-1}\alpha_{2n}}\ .}
Note that the factor ${m(s)/\vert m(s)\vert}= \theta(s)$ for a domain
wall mass, and that this legacy from the heavy modes does not
vanish in the limit $\vert m_0\vert\to \infty$

  Because $\theta(s)$ has opposite signs on the
two sides of the domain wall, the current has a nonzero divergence:
in the presence of nonzero $E$ and $B$ fields, the current has a net
flux onto or off of the domain wall.
  Callan and Harvey pointed out
that this flux precisely accounts for the anomalous divergence \anom\ of the
zeromode current.  Therefore,
what appears as an anomalous divergence in the effective $2n$
dimensional theory is simply charge flowing on or off the mass defect
from the extra dimension.  Note that the sign of the mass coupling in
eq. \lmass\  determines
both the sign of the Goldstone-Wilczek current \gwc\ and the chirality
$\Gamma_5$ of the zeromode on a given mass defect, which figures in the
anomaly  equation \anom. Callan and Harvey also showed that the
 current \gwc\ may be directly computed from an effective action,
which is just the Chern-Simons term  \ref\jack{S. Deser, R. Jackiw and
S. Templeton, \prl{48}{1982}{975}; Ann. Phys. (N.Y.) 140 (1982) 372.}\
proportional to ${m(s)/\vert m(s)\vert}$ that was
induced in the effective theory by integrating out the massive fermion
modes \ref\oddam{N. Redlich, \prl{52}{1984}{18}; A.J. Niemi and G.
Semenoff, \prl{51}{1984}{2077}}\ \foot{See Naculich   \ref\nac{S.G. Naculich,
\np{296}{1988}{837}}\ for a discussion of the subtleties involved. }.

The extra dimension is the loophole in the
Nielsen-Ninomiya theorem \nn\ through which the fermions have wriggled.
The currents of the $2n+1$ dimensional lattice are in fact exactly
conserved. Nevertheless, there is anomalous
current violation in the $2n$ dimensional effective theory  due to
charge flowing in from the extra dimension.  As the
Chern-Simons action is the only new local operator that can be induced in the
effective theory that does not vanish as the fermion mass increases, the
only effect the massive fermions have on the low energy  theory (aside
from renormalizing couplings) is to correctly reproduce the anomalous
Ward identities.

As the Goldstone-Wilczek current \gwc\ does not vanish in a regulated
theory, the $2n$ dimensional anomalous Ward identities may for the first
time be directly measured on the lattice. Such a numerical computation has
recently been performed by
K. Jansen \ref\jansen{K. Jansen, UCSD-PTH/92-18.}, whose results
 show that starting from $2+1$ dimensions, one can measure the $1+1$
 anomalous divergence of a chiral current.  He finds that without the
Wilson term \wact\ the theory is vectorlike and all $2n$ dimensional
flavor currents are conserved. However, for nonzero Wilson coupling $w$,
he finds a nonzero divergence for the currents, and the rate
of charge violation on the domain  wall is proportional to the applied
electric field, as prescribed by the anomaly equation \anom.
Chiral theories with nontrivial anomaly cancellation can also be
simulated in background gauge fields.  The simplest example of such a
theory is the 3-4-5 chiral Schwinger model: one starts with a $2+1$
dimensional theory with three Dirac fermions $\Psi_3$, $\Psi_4$ and
$\Psi_5$ coupled to a $U(1)$ gauge
field with charges $q=3,4$ and 5 respectively.  The mass term for these
fermions is given by
\eqn\mterm{\CL_m = m(s)\left[\bar \Psi_3 \Psi_3 + \bar \Psi_4 \Psi_4 -
\bar \Psi_5 \Psi_5\right]\ , }
so that the effective $1+1$ dimensional theory on the $s=0$ defect
consists of charge
$q=3,4$ fermions with $\Gamma_5=+1$, and a $q=5$ fermion with
$\Gamma_5=-1$.   As  the 2-dimensional
anomaly in the gauge current is proportional to $\Tr q^2\Gamma_5 $, it
vanishes; the Chern-Simons term induced by the heavy fermions is
proportional to $\Tr
q^2 (m/\vert m\vert)$, and so it  and the resultant Goldstone-Wilczek
current also vanish.  As has been confirmed numerically by Jansen
\jansen, the gauge current is anomaly free, even though the individual
flavor currents are anomalously divergent.

The next step is to introduce dynamical gauge fields.
Including the usual $2n+1$ dimensional Yang-Mills action
is not appropriate for describing a $2n$ dimensional gauge on the domain wall.
There would exist an additional polarization for the gauge bosons,
corresponding to the $2n+1$ component of the vector
potential and appearing as an adjoint pseudoscalar in the $2n$ dimensional
world.
Another problem is that the gauge bosons would propagate off  the defect,
giving rise to the wrong force law between the zeromodes.
The conventional $2n+1$ dimensional gauge action is not
required though, since the  action need only be invariant under gauge
and $2n$ dimensional Lorentz transformations.  Thus I consider a gauge
action of the form
\eqn\gaction{
 S_g = \beta_1 \sum_{i,j=1}^{2n}\Tr U^4 \vert_{\lform_{ij}} + 2\beta_2
\sum_{i}\Tr U^4\vert_{ \lform_{i,s}}\ .}
In general,  $\beta_1$ and $\beta_2$ are independent and can
be functions of $s$.  It appears that simply setting to zero the (renormalized)
coupling $\beta_2$ would suffice for simulating a $2n$ dimensional gauge
theory at each $s$-slice.  However, in the
interacting theory, this is clearly not always possible.
It is instructive to consider the continuum
Yang-Mills equations  for these gauge bosons in $2n+1$
dimensions,
\eqn\yme{\eqalign{
\beta_1 D_j F_{ij} +  D_s \beta_2 F_{is} &= J_i\cr
\beta_2 D_i F_{si} = J_s\ .}}
{}From the second equation one sees that $F_{is}\sim J_s/\beta_2$.
Therefore, taking the limit $\beta_2\to 0$ does not eliminate the
unwanted $F_{is}$ term from the first equation---unless $J_s$
vanishes.

There are two possible contributions to $J_s$, from the zeromodes and
from the Goldstone-Wilczek current \gwc.
In fact in the infinite volume limit there is no zeromode contribution to the
to $J_s$ since $\bar \psi \gamma^{2n+1} \psi = \bar \psi
\Gamma_5 \psi=0$, where $\psi(x)$ is any of the zeromodes on  the mass
defect\foot{At finite volume there is an
exponentially small probability for the zeromode to tunnel to the
anti-domain wall, contributing to $J_s$.}. Therefore, for
$J_s$ to vanish,  the component of the Goldstone-Wilczek
current orthogonal to the domain wall must vanish.  As we have seen,
{\it this occurs if, and only if, the gauge currents in the $2n$
dimensional theory are anomaly free}.

I conclude that if the $2n$ dimensional gauge currents are
anomalous, there is nothing one can do to make the theory look like a
$2n$ dimensional gauge theory---it looks inherently $2n+1$ dimensional,
where it makes sense.
However, I speculate  that for anomaly free
chiral gauge currents, the $\beta_2 \to 0$ limit decouples the unwanted
gauge degrees of freedom, and the lattice model looks like a $2n$
dimensional chiral gauge theory.  In that case, 3- and 5-dimensional
models of this nature are expected to possess second order phase
transitions, corresponding to continuum chiral gauge theories in 2- and
4-dimensions.  Note that the effective $2n$ dimensional gauge coupling
will be related to $\beta_1$ by a factor of the  normalization of the
transverse wavefunction $\phi_+(s)$, correctly accounting for the change
in  dimensionality.

Suppose the mechanism described here works and one can simulate the
standard model---how does the
proton decay?  As I argued above, the only requirement for recovering a
4-dimensional gauge symmetry was that the chiral {\it gauge} current be
anomaly-free.  The global currents, such as $B$ and $L$ will still be
anomalous. Given an $SU(2)$ instanton event in the $s=0$ mass defect, a
proton will flow off the domain wall and its baryon number will reside
in the vacuum polarization of the extra dimension.  Since $B-L$ is
anomaly free, the vacuum  will disgorge a positron into
our 4-dimensional world (I am assuming a single family).  As far as we
are concerned, the proton has decayed; little do we suspect that it is
hovering nearby but out of sight in the extra dimension.

\vfill\eject
\centerline{\bf Acknowledgements}
\vskip.5in
I would like to thank T. Banks, A. Cohen,  K. Jansen, J. Kuti,  A.
Manohar, M. Peskin, J. Preskill and S. Sharpe for useful conversations.
After hearing about this work, J. Preskill pointed out to me
that the authors of ref. \ref\frad{D.
Boyanovsky, E. Dagotto and E. Fradkin, \np{285}{1987}{340}; E. Fradkin,
{\it ibid.} (Proc. Suppl.) 1A (1987) 175; E. Dagotto, E. Fradkin and
A. Moreo, \pl{172}{1986}{383}}\ previously discussed domain walls and
 the Callan-Harvey effect on the lattice.  They  mentioned the
possibility of using fermion zeromodes on domain walls to simulate
chiral lattice theories, but rejected it.
\bigskip
\noindent
This research was
supported in part by the Department of Energy under contract
\#DE-FGO3-90ER40546,  by the NSF under contract PHY-9057135,
 and by a fellowship from the Alfred P. Sloan Foundation.

\listrefs

\end